\documentstyle[12pt]{article}
\begin{document}

\begin{center}
{\bf ON THE {\it 1/f} FLUCTUATIONS IN THE NONLINEAR SYSTEMS AFFECTED BY NOISE%
}\\[\baselineskip]

{\ }{\bf B.\,Kaulakys}{\small \ \\Institute of Theoretical Physics and
Astronomy, A. Go\v stauto 12, 2600 Vilnius, Lithuania, \\[\baselineskip]}
{\bf T.\,Me\v skauskas}{\small \ \\Department of Mathematics, Vilnius
University, Naugarduko 24, 2006 Vilnius, Lithuania}\\[\baselineskip]
\end{center}

\vspace{0.2cm}

\begin{center}
ABSTRACT
\end{center}

\hspace{0.5cm} We investigate a problem of the necessary and sufficient
conditions for appearance of the $1/f$ fluctuations in the simple systems
affected by the external random perturbations, i.e. the power spectral
density of the flux of particles moving in some contours and perturbed by
the external forces. In some cases we observe the $1/f^\delta $ $(\delta
\simeq 1)$ behavior but only in same range of frequencies and parameters of
the systems.

\begin{center}
INTRODUCTION
\end{center}

\hspace{0.5cm} Many processes of the greatest diversity exhibit temporal
evolution characterized by a power spectral density $S(f)$ which behaves
like $1/f^\delta $ $(\delta \simeq 1)$ at low frequencies. Despite
considerable efforts, the generally accepted theory of the $1/f$ noise is
still lacking. Possibly, there are several physical mechanisms giving rise
to $1/f$ noise.

\hspace{0.5cm} Usually a study of $1/f$ noise in materials starts from the
approximate empirical relation given in 1969 by Hooge which for the current $%
I$ may be written as [1]
$$
S(f)=\bar I^{{}\,^{\scriptstyle 2}}\frac \alpha {Nf}.\eqno{(1)}
$$
Here $\bar I^{{}\,^{\scriptstyle 2}}$ is the squared average current, $%
\alpha $ is a dimensionless constant which characterizes the noise intensity
and $N$ is the total number of the free charge carriers or fluctuating
particles. Since the current is proportional to the number of particles, the
power spectral density is proportional to the number of particles $N$ too.
Therefore, it is likely that every particle influences on the spectral power
additively and it is logical to search of the $1/f$ fluctuations in the
few-particle systems affected by some perturbations.

\begin{center}
MODELS AND RESULTS
\end{center}

\hspace{0.5cm} One of the relatively universal and approved mechanisms which
generates $1/f$ noise is the intermittency [2]. Intermittent chaos is often
observed phenomenon in nonlinear dynamical systems. Investigating a
transition from chaotic to nonchaotic behavior in the randomly driven
systems [3] we have observed the intermittency route to chaos too [4].
Therefore, it is interesting to analyze the power spectral density of the
flux of small number of particles, or even one particle, moving in some
contours and perturbed by external random forces.

\hspace{0.5cm} The intensity of one-dimensional flux of small particles in
some space point $x_s$ may be expressed as
$$
I(t)=\sum_k\mathop{\rm sign}\nolimits(v_k)\,\delta (t-t_k),\eqno{(2)}
$$
where $t_k$ is a sequence of transit times when one of the particles crosses
the point $x_s$ and $\mathop{\rm sign}\nolimits(v_k)$ is the sign of
velocity in this point. The power spectral density of the flux (2) is
$$
S(f)=\frac 2T\left\langle \left| \sum_k\mathop{\rm sign}\nolimits(v_k)e^{%
\displaystyle -i2\pi ft_k}\right| ^2\right\rangle ,\eqno{(3)}
$$
where $T$ is the whole observation time and the brackets $\left\langle
...\right\rangle $ denote averaging over the realizations. When the averaged
flux $\bar I$ is nonzero the dominant term in the power spectral density at
low frequencies regardless of the detailed structure of the flux is $%
S_{dc}(f)=\bar I^{{}\,^{\scriptstyle 2}}/T(\pi f)^2$. To eliminate this term
we divide the whole observation time $T$ into pieces of equal length and
calculate the averaged power spectral density of the difference of two
pieces fluxes [6].

\hspace{0.5cm} We analyze the power spectral density of particles moving in
some potential and perturbed by the random noise modelled like in papers [3,
4] by the resets of velocity of all particles after every time interval $%
\tau $ according to the replacement $v^{new}(k\tau )=\alpha v^{old}(k\tau
)+\beta v_k^{ran},\quad k=1,2,\ldots $.Here $\alpha $ and $\beta $ are
positive parameters and $v_k^{ran}$ are random velocities from the Gaussian
distribution of variance $\sigma =1$. First of all we have calculated the
power spectral density of the flux of particles bounded in the Duffing
potential $V\left( x\right) =x^4-x^2.$ Under some conditions a transition
from chaotic to nonchaotic behavior and synchronization of particles in such
system may be observed [3-5]. The calculated spectral density of the current
of particles in such systems usually was very close to the white noise. Only
for $\alpha =1$ we observe the $1/f^2$ noise, while for $\alpha $ close to 1
the noise dependence on the frequency is between $1/f^2$ and $1/f$ (see Fig.
1(a)). A very small component of $1/f$ noise in the flux fluctuations of
such system may be explained on the basis of Eq.\thinspace (1) according to
which the intensity of $1/f$ noise component is proportional to the squared
averaged current $\bar I^{{}\,^{\scriptstyle 2}}$. But in such model the
averaged current equals zero.

\hspace{0.5cm} Therefore further we have analyzed the power spectral density
of the current of particles moving in the closed contour and perturbed by
the common for all particles noise. The simplest equations of motion for
such model are of the form [5]
$$
\dot v=F(x)-\gamma v,\qquad \dot x=v\eqno{(4)}
$$
Here $F(x)=F(x+L)$ is the (regular) periodic in the space force with $L$
being the length of the contour and $\gamma $ is the friction coefficient.
Because of the directed drift the averaged current in this model is nonzero.
We have investigated the dependence of the power spectral density on the
nonlinearity of equations of motion and on intensity of the perturbation.
Moreover, we have analyzed the ensemble of particles moving with the
friction in the periodically time-dependent force $F(x+L,t+T_p)=F(x,t)$ and
affected of the identical for all particles noise. We have observed the
relatively weak sensitivity of the spectral density dependence on the
frequency to the nonlinear and periodically time-dependent terms in the
equation for the velocity. Moreover, the spectral density is not sensitive
to the number of the particles $N$ moving in the contour too. Therefore, we
show results of calculations for the simplest versions of the model (4),
i.e. for the motion of one particle affected by the constant force and for
the free motion with perturbations modelled by the velocity replacements
every time interval $\tau $ (see Figs. 1 (b) and (d)). For some parameters $%
\tau $ when $\alpha \neq 1$ we observe the $S(f)$ dependence on the
frequency $f$ (in same range of frequencies) close to the $1/f$-dependence.
Although, when we analyze the current power spectral density in the large
range of frequency we can, as a rule, discover the Lorentzian-type curves.
However the increase of the $S(f)$ with the decrease of the frequency
usually is not like $1/f^2$ but often as $1/f$.

From the analyzes above it appears that the current power spectral density
in the most degree depends not on the type of the potential (and force $%
F(x,t)$) in which the particles move but much more on the type and intensity
of the perturbation. Therefore, we have also analyzed the simplest model for
the perturbed flux: the free motion of the particles with the Brownian-type
additional fluctuations of the transition times. For such model the transit
times may be calculated from the iterative sequence $t_k=t_{k-1}+1+\beta
\delta t_k^{ran}$. Here the period of the free motion in the closed contour
is chosen to be equal 1 and $\delta t_k^{ran}$ are random time increments
from the Gaussian distribution of variance $\sigma =1$. Results of
calculations of $S(f)$ for different values of intensity of perturbation $%
\beta $ are presented in Fig. 1 (c). With increase of the perturbation
intensity we observe transition from the white noise at $\beta \leq 1$ to
the Lorentzian-type, $1/f$- and $1/f^{1.3}$-type noise at $\beta \geq 2$.

\hspace{0.5cm} Summarizing, we have analyzed the problem of necessary and
sufficient conditions for appearance of the $1/f$-type fluctuations in the
current of the simple systems of particles moving in some contour. Even in
such simplest models consisting of few or even one particle moving in the
closed contours and perturbed by the appropriate external noise we may
observe the power spectral density $S(f)$, which behaves like $1/f^\delta $ $%
(\delta \simeq 1)$ in same range of frequencies.

\begin{center}
ACKNOWLEDGMENTS
\end{center}

\hspace{0.5cm} The research described in this publication was supported in
part by the Lithuanian State Science and Studies Foundation.

\begin{center}
REFERENCES
\end{center}

\noindent
1.~F.\,N.\,Hooge, T.\,G.\,M.\,Kleinpenning and L.\,K.\,J.\,Vandamme. {\it %
Rep. Prog. Phys.} \phantom{1. }{\bf 44}, 479 (1981).

\noindent
2.~H.\,G.\,Schuster. {\it Deterministic chaos: an introduction.} (VCH,
Weinheim, 1989).

\noindent
3.~B.\,Kaulakys and G.\,Vektaris. {\it Phys. Rev. E} {\bf \ 52}, 2091 (1995).

\noindent
4.~B.\,Kaulakys and G.\,Vektaris. In: {\it Proc. 13th Int. Conf. on Noise in
Physical \phantom{1. }Systems and 1/f Fluctuations.} Eds V.\,Bareikis and
R.\,Katilius. (World Scientific: \phantom{1. }Singapore, 1995) p.677.

\noindent
5.~B.\,Kaulakys, F.\,Ivanauskas and T.\,Me\v skauskas. In: {\it Proc.
Intern. Conf. on \phantom{1. }Nonlinearity, Bifurcation and Chaos: the Doors
to the Future.} (Lodz-Dobieszkow, \phantom{1. }Poland, September 16-18,
1996) p.145.

\noindent
6.~T.\,Musha and H.\,Higuchi. {\it Jap. J. Appl. Phys.} {\bf 15}, 1271
(1976).

\end{document}